\documentclass[prb,twocolumn,superscriptaddress,floats,showpacs]{revtex4}

\usepackage{graphicx}
\usepackage{dcolumn}
\usepackage{bm}

\begin{document}
\title{Magnetic excitations of the charge stripe electrons below half doping in La$_{2-x}$Sr$_{x}$NiO$_4$, $x = 0.45$ and $x = 0.4$ }
\author{P. G. Freeman}
\affiliation{Jeremiah Horrocks Institute for Mathematics, Physics, and Astronomy, University of Central Lancashire, Preston, PR1 2HE, UK}
\affiliation{Laboratory for Quantum Magnetism, Ecole Polytechnique F\'{e}d\'{e}rale de Lausanne (EPFL), CH/1015 Lausanne, Switzerland }
\affiliation{Helmholtz-Zentrum Berlin, Hahn-Meitner-Platz 1, 14109 Berlin, Germany}
\affiliation{Institute Laue-Langevin, BP 156,
38042, Grenoble Cedex 9, France.}\email{pgfreeman@uclan.ac.uk}
\author{S. R. Giblin}
\affiliation{ISIS Facility, Rutherford Appleton Laboratory, Chilton,
Didcot, Oxon, OX11 0QX, UK} 
\affiliation{School of Physics and Astronomy, Queens Building, The Parade, Cardiff University, CF24 3AA, UK} 
\author{K. Hradil}
\affiliation{Institut f\"{u}r Physikalische Chemie, Universit\"{a}t
G\"{o}ttingen, Tammanstrasse 6, 37077 G\"{o}ttingen, Germany}
\affiliation{T.U Wien, X-ray Centre, Getreidemarkt 9,  A-1060 Vienna, Austria}
\author{R. A. Mole}
\affiliation{Forschungs-Netutronquelle Heinz Maier-Leibnitz (FRM II), Lichtenbergstra\ss e, 1 85747 Garching, Germany}
\affiliation{Bragg Institute, ANSTO, New Illawarra Road, Lucas Heights, NSW,  Australia}
\author{P. Cermak}
\affiliation{J\"{u}lich Centre for Neutron Science at Heinz Maier-Leibnitz Zentrum (MLZ),  Forschungszentrum J\"{u}lich GmbH, Lichtenbergstra\ss e, 1 85748 Garching,
Germany}
\author{D. Prabhakaran 
}
\affiliation{Department of Physics, Oxford University, Oxford, OX1
3PU, UK}

\begin{abstract}
The low energy magnetic excitation spectrum of charge stripe ordered La$_{2-x}$Sr$_{x}$NiO$_4$, $x = 0.4$ and $x = 0.45$ samples were studied by neutron scattering.
 Two excitation modes are observed in both materials, one from the ordered magnetic moments, and a second mode consistent with pseudo-one-dimensional antiferromagnetic excitations of the charge stripe electrons (q-1D). The dispersion of the q-1D excitation follows the same relation as in $x = 1/3$ composition, with even spectral weight in the two counter-propagating branches of the $x = 0.4$ sample, however in the $x = 0.45$ sample only one dispersion branch has any measurable spectral weight. The evolution of the q-1D excitations on doping to the checkerboard charge ordered phase is discussed.

\end{abstract}

\maketitle


Variation between the hourglass shaped magnetic excitation spectrum of the cuprates  compared to the magnetic excitations of charge-stripe ordered La$_ {2-x}$Sr$_ {x}$NiO$_ {4+\delta}$ (LSNO) has called into question the relevance of the charge-stripe picture for the 
cuprates.\cite{tranquada-Nature-1995,hourglass,boothroyd-PRB-2003,bourges-PRL-2003,Woo}  Recently  however, it has  been shown that insulating charge-stripe ordered La$_ {5/3}$Sr$_ {1/3}$CoO$_ {4+\delta}$ (LSCoO) has an hourglass shaped magnetic excitation spectrum that can be explained within a similar linear spin wave model used to describe charge-stripe ordered LSNO.\cite{boothroyd-Nature}
In the charge-stripe model the reason behind the different excitation spectrums is in part due to disorder, but mainly due to the relative strength of the  inter-stripe magnetic interaction across the charge-stripe, relative to the the intra-stripe interaction.\cite{boothroyd-Nature} For a weak inter-stripe interaction an hourglass excitation is observed, whereas a strong interaction leads to the symmetric spin wave cones observed in LSNO.\cite{boothroyd-PRB-2003,Woo} The role of disorder in wiping out the mode that disperses away from the antiferromagnetic position has been established in a charge-stripe ordered manganate, these modes are present at base temperature but they have no measurable intensity on warming to the magnetic ordering temperature.\cite{Ulbrich.arxiv}  In the case of La$_ {2-x}$Sr$_ {x}$CoO$_ {4+\delta}$ an alternative scenario for the hourglass excitation spectrum arising from disordered checkerboard charge-ordered state has been proposed.\cite{dress-NatComm-2013} With the recent direct evidence of charge-stripe order in La$_ {2-x}$Sr$_ {x}$CoO$_ {4+\delta}$, the origin of the magnetic hourglass excitation spectrum is in dispute in this material.\cite{Babkevich_NatComm_2016}

\begin{figure}
\includegraphics[width=8.4cm]{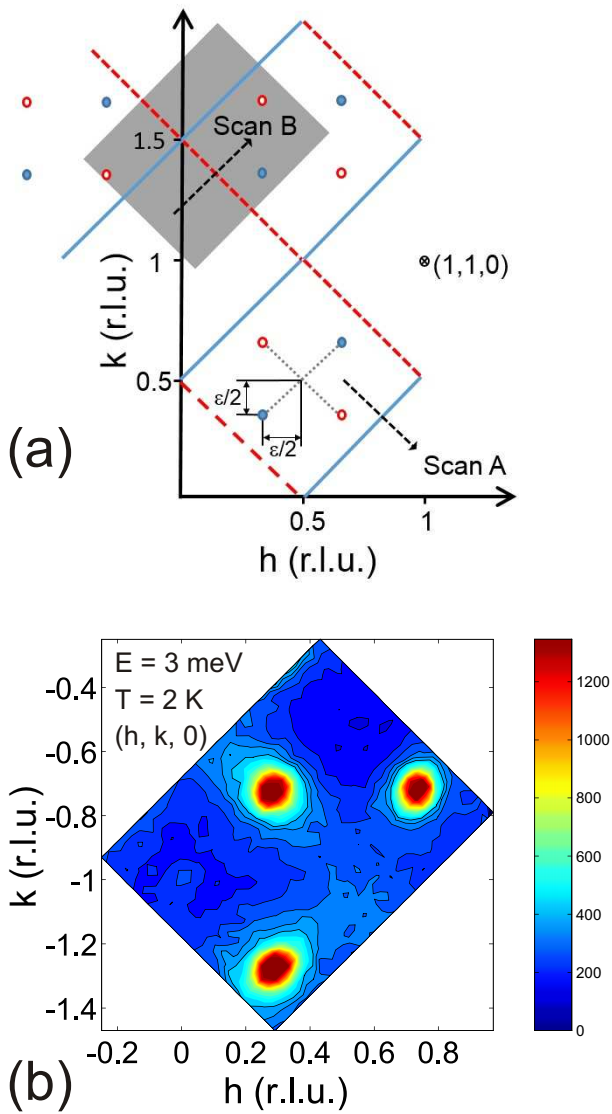}
\caption{(Color online) (a) A diagram of the $(h, k, 0)$ plane in reciprocal space for La$_ {2-x}$Sr$_ {x}$NiO$_ {4+\delta}$. The location of magnetic Bragg reflections of the two spin stripe domains are represented by circles, and the $(1, 1, 0)$  Bragg reflection is indicated. Dashed and solid lines indicate the location of gapped quasi-one-dimensional antiferromagnetic correlations of the charge-stripe electrons (q-1D) at $\approx 1.5$\,meV. The filled circles and solid lines correspond to $[1, -1, 0]$ charge-stripe domains, and open circles and broken lines are for $[1, 1, 0]$ stripe domains. The dotted arrows indicate the paths of two scans used to study the q-1D excitation in this paper, whereas the grey shade rectangle is the approximate location of the grid scan used to study La$_{2-x}$Sr$_{x}$NiO$_4$ $x = 0.45$ at 3\,meV shown in Fig.\ref{2Dmapq1d}. (b) Low-energy scattering from La$_ {3/2}$Sr$_ {1/2}$NiO$_ {4}$ measured by a grid of scans at 3\, meV.
The three centres of scattering are from excitations from the spin-stripe ordered Ni$^{2+}$ spins, and the extended ridges underneath these excitations are from quasi-one dimensional antiferromagnetic excitations of the charge stripe electrons. The existence of two modes  was confirmed by the temperature dependences of the excitations.\cite{freeman-PRB-2005}} \label{halfdoped}
\end{figure}

In LSNO magnetic excitations of the charge-stripe electrons are observed,\cite{boothroyd-PRL-2003} whereas half-filled charge-stripes in the cuprates appear to have no inter unit cell spin interactions, and 
the spin state of the charge stripes in LSCoO  is effectively $S = 0$.\cite{tranquada-Nature-1995,hourglass} In the  case of the cuprates there is evidence for a $\bf{Q} = 0$ intra unit cell magnetic order from certain experimental techniques,\cite{fauque-PRL-2006}, this is apparently effected by charge-stripe order in La-based cuprates possibly indicating the involvement of spins of half-filed stripes.\cite{baledent_PRL_2010}  It is necessary to understand the magnetic interactions of the charge-stripe electrons in LSNO to throw light on how they mediate the strong inter-stripe spin interaction in these materials, which results in the spin wave cone dispersion of the magnetic excitations.  

It has been shown in LSNO around $x \sim 1/3$  there are doping independent gapped quasi-one-dimensional antiferromagnetic correlations along the charge-stripes of the $S = 1/2$ charge-stripe electrons (q-1D).\cite{boothroyd-PRL-2003,freeman-PRB-2011} Where the periodicity of the order is four Ni sites long. Figure \ref{halfdoped}(a) shows the wave vector positions of magnetic excitations  from the ordered  Ni$^ {2+}$ $S = 1$ and the magnetic excitations form the q-1D mode in LSNO.  In the  $x = 0.5$ the charge order is part stripe, and part checkerboard in character.  The q-1D excitation in the $x= 0.5$ was observed to lock into the wavevector of the ordered Ni$^ {2+}$ $S = 1$ spins at  $(h + 1/2 \pm \varepsilon /2, k + 1/2 \pm \varepsilon /2, l)$, $\varepsilon = 0.445$ positions in reciprocal space, as shown in Fig. \ref{halfdoped}.\cite{freeman-PRB-2005} In LSNO $\varepsilon$ is known as the incomensurability, where $1/\varepsilon$ is the average periodicity of the charge order. Further details of the low energy magnetic excitations of the $x = 0.50$ material such as bandwidth and zone boundary of the q-1D excitation, remain unclear due to the broadening of magnetic excitations  in comparison to the  $x = 1/3$ material.\cite{freeman-PRB-2005} In this paper the issue of the q-1D excitation's lock-in in the $x = 0.5$ composition is addressed by studying the striking difference of the q-1D excitation between $x = 1/3$ and $x = 0.5$ doping.


Single crystals of La$_{2-x}$Sr$_{x}$NiO$_4$ $x = 0.4$ and $x = 0.45$ compositions were grown using the floating-zone technique.\cite{prab}. The $x = 0.4$   sample was a slab of dimensions $\approx 15 \times 10 \times 4$\,mm and weighs 1.8\,g. The  $x = 0.45$ sample investigated was a rod of 6\,mm in diameter and 25\, mm in length  weighing  2.5\,g. The samples used here are the same samples that were studied in previous neutron diffraction measurements.\cite{freeman-PRB-2004,giblin-PRB-2008}  Oxygen content of as grown $x = 0.4$ sample was determined to be stoichiometric by thermogravimetric analysis,\cite{prab}  and the results of the neutron diffraction study of  $x = 0.45$  sample are consistent with stoichiometric oxygen content.\cite{giblin-PRB-2008,yoshizawa-PRB-2000} The bulk magnetization of  the $x = 0.45$ sample is found to have the same characteristics as a slightly oxygen deficient  $x = 0.5$ sample, but the $ x = 0.45$ sample has the advantage of being well removed from the checkerboard charge ordered state.\cite{giblin-PRB-2008,freeman-JSNM-2011,kajimoto-PRB-2003}

Neutron scattering experiments were performed on the triple-axis spectrometers PUMA,\cite{puma} and PANDA\cite{panda} at the Heinz-Maier Leibnitz Zentrum, and IN8 at the Institut Laue-Langevin.\cite{in8} The data were collected with a fixed final neutron wavevector of $k_f = 2.662$\,\AA\  on PUMA and IN8, and $k_f = 1.55$\,\AA\  on PANDA. On PUMA and IN8 a pyrolytic graphite (PG) filter and on PANDA a Beryllium filter, was placed after the sample to suppress higher-order harmonic scattering. The excitation spectrum of  the  $x = 0.45$ sample was measured on PUMA and PANDA, and the excitation spectrum of the $x = 0.4$ sample was measured on IN8.  On all instruments the neutrons final and initial energy was selected by Bragg reflection off a double focusing pyrolytic graphite (PG) monochromator and double focusing analyzer (PUMA and IN8 and PANDA).  The sample was orientated so that on all instruments  ($h$,\ $k$,\ 0) positions in reciprocal space could be accessed. In this work the tetragonal unit cell of LSNO is referred to, with unit  cell parameters $a \approx 3.8$\,\AA\ , $c \approx 12.7$ \AA. A report on the magnetic excitation spectrum from the ordered Ni$^{2+}$ spins has been reported elsewhere,\cite{freeman-JPKS-2013} with the excitation spectrum extending to $> 50$\,meV.

\begin{figure}
\includegraphics[width=8.4cm]{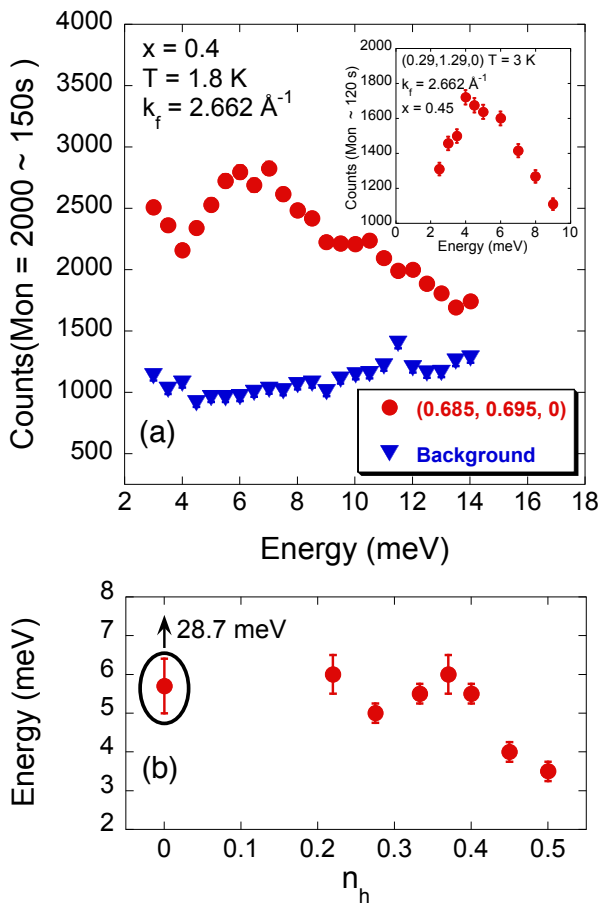}
\caption{(Color online) (a) An energy scan of the magnetic excitations at the magnetic zone centre of La$_{2-x}$Sr$_{x}$NiO$_4$, $x = 0.4$ sample. The background was measured at a wave vector  $\pm (0.15, 0.15,0)$ stepped away from the magnetic zone centre, sufficiently spaced to avoid the q-1D excitation. Inset: An energy scan of the magnetic excitations at the magnetic zone centre of $x = 0.45$ sample. In both energy scans the magnetic intensity first increases as the out-of-plane anisotropy gap is over come, then decreases with increasing energy transfer as expected for antiferromagnetic excitations. (b) The doping variation of the out-of-plane anisotropy gap of La$_ {2-x}$Sr$_ {x}$NiO$_ {4+\delta}$ where $n_h = x + 2\delta$ is the hole concentration. Additional data points taken from previous studies.\cite{Woo,freeman-PRB-2005,Nakajima-JPSJ-1993,boothroyd-PhysicaB-2004,freeman-PRB-2009}}
\label{anisoptropy}
\end{figure}
	
In the energy range of  this study there is little development of the dispersion of magnetic excitations from the $S = 1$  Ni$^{2+}$ spins, with inelastic neutron scattering unable to resolve the counter propagating modes. In Fig. \ref{anisoptropy}(a) an energy scan of the magnetic excitations of the $x = 0.4$ sample at the magnetic zone centre with, $\varepsilon = 0.37$ is shown. With increasing energy transfer the intensity is approximately constant before increasing in-between 4 and 5.5\,meV, then decreasing in intensity monotonically above 7\,meV. The inset of Fig. \ref{anisoptropy}(a) shows that a similar behaviour is observed for the $x = 0.45$ sample at the magnetic zone centre with $\varepsilon = 0.42$. For the $x = 1/3$ material and the a $x = 0.5$ material polarized neutron scattering determined that this increase in intensity at low energy transfer is due to overcoming the out-of-plane anisotropy gap in LSNO, which is greatly reduced compared to the parent material.\cite{boothroyd-PhysicaB-2004,freeman-PRB-2005} We therefore assume that the increase intensity observed at similar energies in other doping levels is the out-of-plane anisotropy, including here for the $x = 0.4$ and $x = 0.45$ samples. In Fig. 2(b) the doping dependence of the assumed out-of-plane anisotropy gap is shown, with the anistropy gap energy defined to be the lowest energy at which maximum intensity occurs. We note that the reduction in the out-of-plane anisotropy gap of charge-stripe ordered LSNO is similar to that reported for La$_{2-x}$Sr$_{x}$CuO$_4$.\cite{romer-PRB-2013}	


\begin{figure}\includegraphics[width=8.4cm]{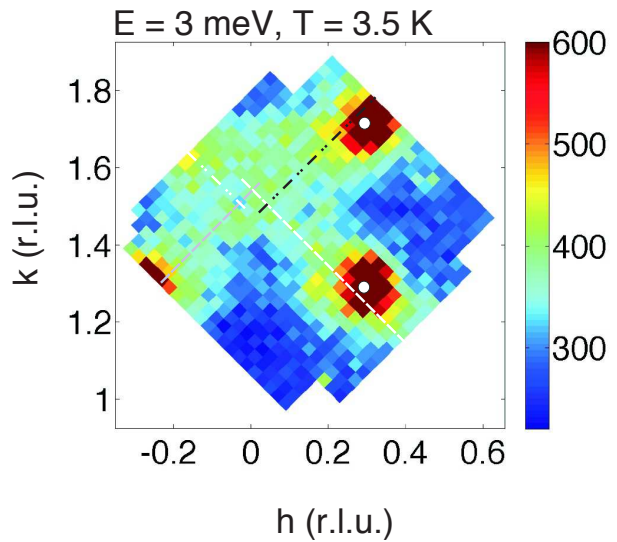} \caption{(Color online) A grid of low energy neutrons scans on the $x = 0.45$ sample perfromed at 3  meV, performed on PUMA. White circles indicate the wavevector of Bragg reflections from the spin-stripe ordered Ni$^{2+}$ spins, whilst the four dashed lines are guides to the eye of the four diffuse ridges of scattering observed in this grid scan.}
 \label{2Dmapq1d} \end{figure}

\begin{figure}\includegraphics[width=8cm]{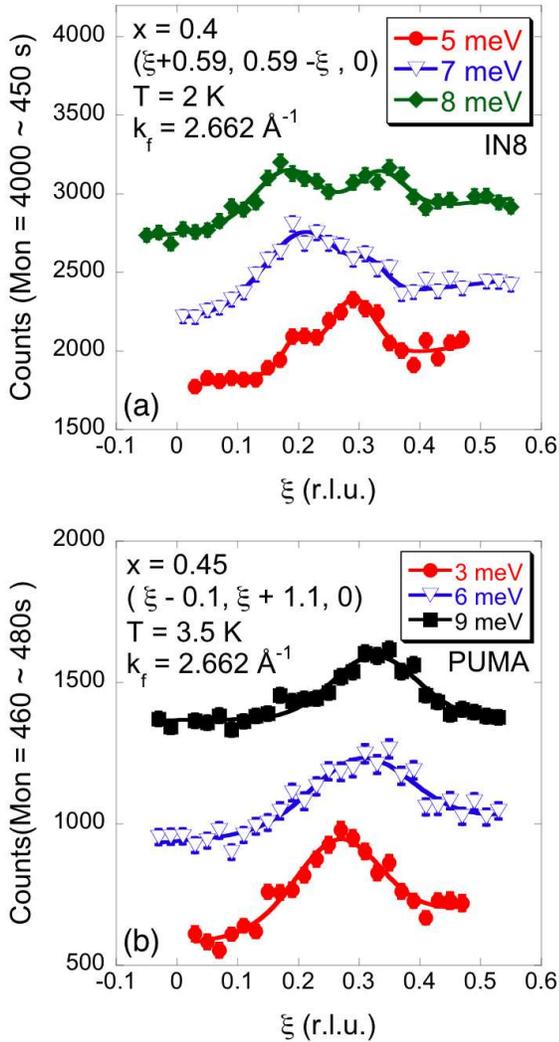} .\caption{ (color online) Constant-energy scans across the q-1D excitation, 
for (a) the $x = 0.4$ sample (scan A), (b) the $x = 0.45$ sample (scan B). Successive scans have been offset vertically by the addition of (a) 500 counts (b) 250 counts, for the purpose of clarity. The solid lines are the result of a fit to the data of a sloping background with (a) two Gaussian lineshape, and (b) one Gaussian lineshape.} \label{ConstEscans}
\end{figure}

\begin{figure}\includegraphics[width=8cm]{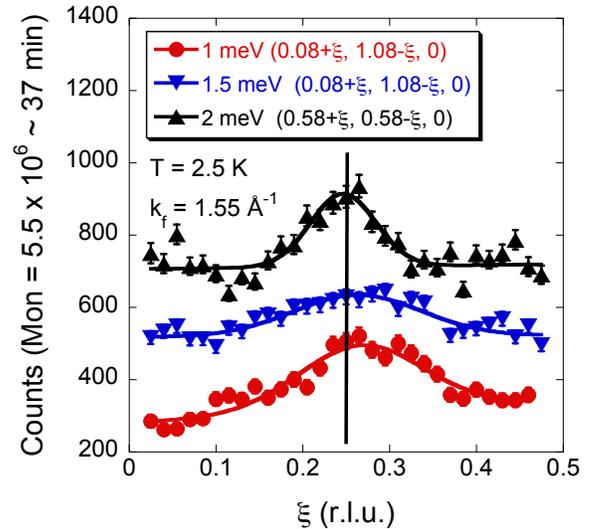} \caption{ (Color online) Constant-energy scans across the q-1D excitation in the $x = 0.45$ sample measured at 1.0\,meV, 1.5\,meV (scan B) and 2\,meV (scan A).  The solid lines are the result of a fit to the data of a sloping background with one Gaussian lineshape. The wave vector resolution parallel to the scan resolution for is 0.0124 r.l.u. at  1.0\,meV,  0.0127 r.l.u. at 1.5\,meV and 0.0131 r.l.u. at 2\,meV. A broadening of the q-1D excitation is observed in scans at 1.0\,meV and 1.5\,meV, compared to 2\,meV.} \label{nonexist}
\end{figure}

\begin{figure}\includegraphics[width=8cm]{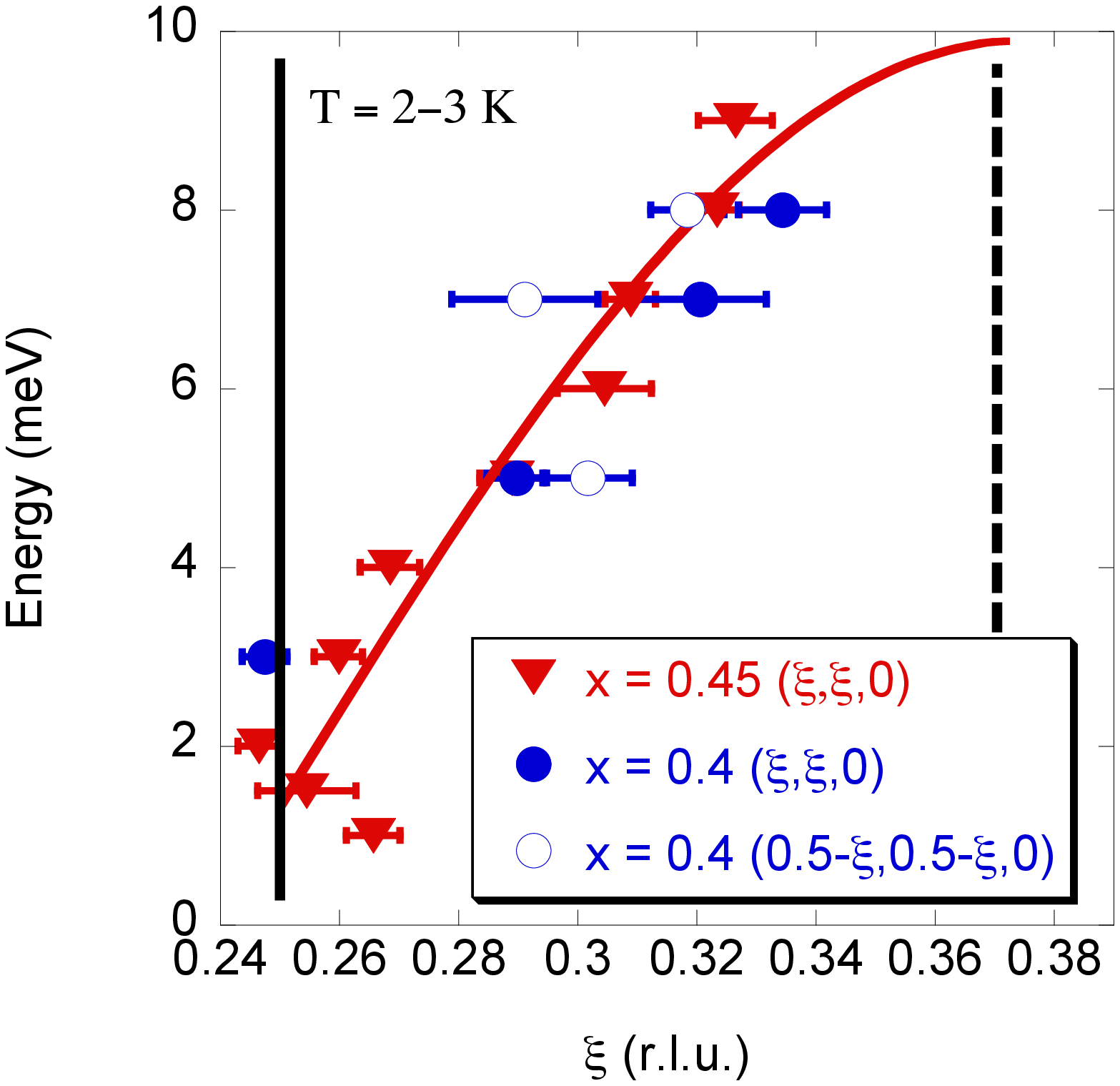} \caption{ (Color online) The fitted centres of the q-1D excitations obtained from constant energy scans such as those shown in Fig \ref{ConstEscans} and Fig.\ref{nonexist}. A solid vertical line represents the zone centre, and a dashed vertical line represents the zone boundary of the q-1D excitation. Centres of the q-1D excitation at $5,7$ and 8\,meV  in the $x = 0.4$ that disperse to smaller values of $\xi$ have been folded into this half Brillioun zone for better comparison with the dispersion of the $x = 0.45$ sample. The $x = 0.45$ sample data is fitted with a gapped sinsusoidal dispersion for the expected q-1D excitation periodicity, with the fit being consistent with the dispersion of the q-1D excitation observed in the $x = 1/3$ material.\cite{boothroyd-PRL-2003,freeman-PRB-2011}
} \label{dispersion}
\end{figure}

In previous studies it has been demonstrated that an effective way to determine the location of the q-1D excitation is to map out reciprocal space in a grid scan at a fixed energy transfer.\cite{boothroyd-PRL-2003,freeman-PRB-2005}  In Fig. \ref{2Dmapq1d} a grid scan of the excitations of the $x = 0.45$ sample at 3\,meV measured at 3.5 K is shown. White circles have been added to this figure to indicate the magnetic zone centres of the ordered Ni$^ {2+}$ spins, where intense magnetic excitations are observed to be dispersing out of the magnetic zone centres. Additional to the zone centre magnetic excitations, diffuse ridges of scattering are present that are consistent with the presence of q-1D correlations.\cite{boothroyd-PRL-2003,freeman-PRB-2005,freeman-PRB-2011}

By performing scans parallel to the q-1D excitations, that pass through the excitations of the ordered Ni$^ {2+}$ spins, we can determine a position where there is zero intensity from the excitations from the ordered spins. At this position scans across the q-1D excitations can be performed to map out the dispersion of the q-1D scattering. Fig. \ref{ConstEscans} shows constant energy scans across the q-1D excitations for the (a) $x = 0.4$ sample and (b) $x = 0.45$ sample. In the $x = 0.4$ sample the constant energy scans are well described by fitting two Gaussian lineshapes on a sloping background, with the splitting of the two peaks increasing with increasing energy transfer. 
Fig. \ref{ConstEscans} (b) however shows that in the $x = 0.45$ sample the dispersion of the q-1D excitations can be fitted by one Gaussian lineshape on a sloping background, that disperses with increasing energy transfer. There is a small shoulder of scattering on the left hand side of the 9\,meV peak that is not perfectly described by this lineshape, and there maybe a relatively weak excitation mode.  The dispersion of the q-1D excitations is clearly different in the two doping levels studied here, in the $x = 0.4$ sample there are two counter-propgating modes, whereas in the the $x = 0.45$ sample only one dispersive mode is observed to have significant spectral weight.


Further measurements of the q-1D excitations have been performed on the $ x = 0.45$ sample to lower energies using the higher energy resolution of the cold triple axis spectrometer PANDA with $k_f = 1.55$\,\AA\ .  In Fig. \ref{nonexist} constant energy scans of the $x = 0.45$ sample at 1.0,\ 1.5\ and 2\,meV are fitted by a Gaussian lineshape on a sloping background. 
The scan of the q-1D excitations at 1.0\,meV and at 1.5\,meV are significantly broadened compared to the scan at 2.0\,meV. Calculations of the resolution ellipsoid of the experimental setup indicate this broadening is not due to focusing and de-focusing effects, the broadening is intrinsic to the q-1D excitations. At  2.0\,meV and 1.5\,meV the peak centres are consistent with a $\xi = 0.25$ centring of the q-1D. In comparison to the intensity of q-1D  excitations at  2.0\,meV and 1.5\,meV there is an apparent $64\pm32,\%$ increase in intensity observed at 1.0,meV. Attempts to measure the energy gap in a higher resolution mode with $k_f = 1.2$\,\AA\  suffered from  low count rates.

To compare the dispersion of the q-1D excitations in the $x = 0.45$ and $x = 0.4$ samples the energy variation of the fitted peak centres obtained in constant energy scans is shown in Fig. \ref{dispersion}. The centres of the q-1D excitations in the $x = 0.4$ sample have been folded into a half Brillioun zone for comparison purposes. 
Within the experimental error there is no significant difference in the dispersion of the q-1D excitations of the  $x = 0.45$ and $x = 0.4$ samples, indicating the main interactions of the q-1D excitations are the same in both materials.

In the $x = 1/3$ the dispersion of the q-1D  excitations was originally compared to a Heisenberg spin one half antiferromagnet spin chain, before the spin gap was determined.\cite{boothroyd-PRL-2003} We  therefore compare the $x = 0.45$ sample dispersion of Fig. \ref{ConstEscans} to that of a Heisenberg spin one half antiferromagnet spin chain with the inclusion of a gap term:
\begin{eqnarray}
E(\bf {Q}) = E_0 + \pi J | sin(2\pi\bf {Q}. \bf{d})|,
 \label{q-1Ddispersion}
\end{eqnarray}
where $E_0$ the spin gap, $J$ is the nearest neighbour exchange interaction per spin,  and $\bf{d}) = [1,1, 0]$ the direction parallel to the charge-stripes. From this fit we
obtain an energy gap $E_0$ = 1.3$\pm$ 0.5 \,meV and a spin exchange interaction per spin of $J = 2.7\pm0.3$\,meV. In the $x = 1/3$ material the energy width of the q-1D excitations was determined by scanning the zone centre excitations to be to be $1.57 \pm 0.17$,meV, indicating the excitations are short lived.\cite{freeman-PRB-2011} If the q-1D excitations in the $x = 0.45$ sample had a $1.6$,meV energy width, this would explain why the q-1D excitations are observed at 1,meV in the $x = 0.45$ sample. The wavevector of the q-1D excitation indicates a four spin object, this object may better map on to a one dimensional Haldane spin chain of integer spins, with a Haldane spin chain having a spin gap and bandwidth of $4 J$ instead of $\pi J$ for a Heisenberg spin one half antiferromagnet spin chain.\cite{Haldane} The recent development of polarized neutron scattering for time-of-flight inelastic spectrometers would provide the ability to separate phonon and magnetic scattering enabling mapping of the magnetic excitations in the 10-20\, meV energy range,\cite{ToFpol} this would enable measurements to determine if the q-1D has continuum of excitations to differentiate between these two situations. Surprisingly the parameters obtained from this phenomenological fit accurately describe the dispersion the q-1D excitation of the $x = 1/3$,\cite{boothroyd-PRL-2003,freeman-PRB-2011} further showing the doping independence of the main  spin interaction of the q-1D excitations.

Charge-stripes in LSNO have been determined to be predominantly centred on Ni sites at low temperature,\cite{wochner-PRB-1998} so that between a third and half doping  the charge-stripe structure is believed to be an admixture of the $\varepsilon = 0.5$ and $\varepsilon = 1/3$ structures .\cite{yoshizawa-PRB-2000} As the doping level is increased towards half doping, the charge-stripe structure becomes closer to the $\varepsilon = 0.5$ of ideal checkerboard charge order. In checkerboard charge order there is simultaneously charge order at 45$^ {\circ}$ to the Ni-O bonds along both diagonal directions of the Ni-O plane. Previously it has been shown that the charge-stripe structure is determined predominantly by the Coulomb interaction between the charge-stripe electrons.\cite{wochner-PRB-1998} The distortion of the lattice by the charge-stripe order is what enables neutron diffraction to study the charge-stripe order.  On going towards the checkerboard charge ordered phase the charge order loses itÕs definition, effectively having zero correlation length out of the Ni-O planes at half doping.\cite{yoshizawa-PRB-2000,kajimoto-PRB-2003} With a reduction in the distortion of the lattice caused by charge order, different pathways for magnetic interactions may therefore be enabled, or enhanced. 

For Ni centred charge-stripe order with  $\varepsilon >  0.418 = (0.5+0.333)/2$, on average a charge-stripe will have one neighbouring charge-stripe 2 Ni-Ni spacings apart as in checkerboard charge order, and another charge-stripe neighbour 3 Ni-Ni spacings apart. For the 2 Ni-Ni spaced charge-stripes there is locally a checkerboard structure, so the charge-stripe electrons will want to interact antiferromagnetically at right angles to the charge-stripe direction. The charge order at right angles to the charge-stripe direction is however too short for the 4 Ni spin object of the q-1D excitation to form, nevermind propagate, so an inter-charge-stripe spin interaction will be limited to a perturbation of the q-1D excitation. Below $\varepsilon = 0.418$ there will be charge-stripes that have two neighbouring charge-stripes 3 Ni-Ni spacings apart, leading to a well-defined stripe direction, and what should be a stronger lattice distortion caused by the charge order that weakens any inter-charge-stripe spin interaction. We tentatively propose that the observed loss in spectral weight of the q-1D excitation dispersing away from the antiferromagnetic position in reciprocal space in the $x = 0.45$ sample is caused by a reduction in the charge-order distortion, that enables a perturbing inter-charge-stripe antiferromagnetic interaction between charge-stripes 2 Ni-Ni spacings apart.

In the cuprates and cobaltates it has been proposed that the hourglass magnetic excitation spectrum arises due a small ratio of interstripe to intrastripe magnetic interactions, and disorder of the spin interactions. \cite{boothroyd-Nature,Ulbrich.arxiv}  The inter-charge-stripe spin interaction we tentatively propose for the loss in spectral weight of one mode of the q-1D excitation in the LSNO $x = 0.45$ material, has a similar role as disorder in producing an hourglass shaped excitation spectrum. We note that the alternative nano-phased separated model of the cobaltates also includes disorder of the magnetic order. \cite{dress-NatComm-2013}

The proposed cause of the change of the q-1D excitations  the $x = 0.45$ material compared to the $x = 0.4$ material is different to the change in the magnetic excitations from the ordered $S = 1$  Ni$^{2+}$ spin-stripes in these two materials.\cite{freeman-JPKS-2013} For the magnetic excitations from the ordered spins it was proposed that the variation in charge-stripe periodicity caused damping of the magnetic excitations from the ordered spins below $\varepsilon = 0.418$, whereas as above $\varepsilon = 0.418$ variation in charge-stripe periodicity is abrupt and causes additional magnetic excitation modes. 

This study has shown the remarkable doping independence of the dispersion of q-1D  magnetic excitations and their quasi-one dimensionality in LSNO over the doping range $x = 0.275$ to $x = 0.45$. Despite the lack of variation in the dispersion of the q-1D scattering there is a dramatic loss of spectral weight in one dispersion branch on going from $ x = 0.4$ to $x = 0.45$ hole doping as the stripe direction definition is reduced. This leads to an alternative explanation to a wavevector lock in of the q-1D to order spin-stripe wave vector in the $x = 0.5$, is an accidental observation of co-incidence due to the softening of the dispersion of the q-1D excitations.\cite{freeman-PRB-2005} Further investigation of the dispersion relation of q-1D excitations in the $x = 0.5$ are warranted.



We wish to acknowledge that crystal growth for this  work was supported by
the Engineering and Physical Sciences Research Council of Great Britain. We wish to acknowledge that this research project has been supported by the European Commission under the 6th Framework Programme through the Key Action: Strengthening the European Research Area, Research Infrastructures. Contract n¡: RII3-CT-2003-505925.


\end{document}